\documentclass[english]{cccconf}
\usepackage{amsfonts}
\usepackage[comma,numbers,square,sort&compress]{natbib}
\usepackage{amsmath}
\usepackage{amssymb}
\begin{document}

\title{Distributed Detection via Bayesian Updates and Consensus}
\author{LIU Qipeng\aref{QDU},
        ZHAO Jiuhua\aref{SJTU},
        WANG Xiaofan\aref{SJTU}}



\affiliation[QDU]{Institute of Complexity Science, Qingdao University, Qingdao 266071, P.~R.~China
        \email{qipengliu@qdu.edu.cn}}
\affiliation[SJTU]{Department of Automation, Shanghai Jiao Tong University, and \\Key Laboratory of System Control and Information Processing, Ministry of Education of China, Shanghai 200240, P.~R.~China
        \email{\{jiuhuadandan,xfwang\}@sjtu.edu.cn}}

\maketitle

\begin{abstract}
In this paper, we discuss a class of distributed detection algorithms which can be viewed as implementations of Bayes' law in distributed settings. Some of the algorithms are proposed in the literature most recently, and others are first developed in this paper. The common feature of these algorithms is that they all combine (i) certain kinds of consensus protocols with (ii) Bayesian updates. They are different mainly in the aspect of the type of consensus protocol and the order of the two operations. After discussing their similarities and differences, we compare these distributed algorithms by numerical examples. We focus on the rate at which these algorithms detect the underlying true state of an object. We find that (a) The algorithms with consensus via geometric average is more efficient than that via arithmetic average; (b) The order of consensus aggregation and Bayesian update does not apparently influence the performance of the algorithms; (c) The existence of communication delay dramatically slows down the rate of convergence; (d) More communication between agents with different signal structures improves the rate of convergence.
\end{abstract}


\footnotetext{This work was supported by the National Natural Science Foundation of China under Grant Nos. 61374176 and 61473189, and the Science Fund for Creative Research Groups of the National Natural Science Foundation of China (No. 61221003).}

\section{Introduction}

Recent years have witnessed a considerable amount of work on analysis of networked systems ranging from social and economic networks to robot and sensor networks \cite{jackson,ren,surveillance,boyd}. An amazing phenomenon arising in networked systems is that, by communicating and cooperating among individuals, the whole group could complete very complicated tasks way beyond the ability of any single agent.

A common task in networked systems is that all agents are supposed to collectively find out an underlying true state of an object using relatively local information such as private observations and neighbors' information. For instance, voters attempt to find out the ability of some political candidates; costumers learn the quality of a new product; a network of sensors detect the mean temperature of a wide area. According to specific contexts, this task might have different names, such as social learning, distributed detection, distributed estimation, and distributed hypothesis testing \cite{learning,deron,rao,jad_detect,olfati,nedic,lalita,UCLA,jad_nb,tahbaz}. To be consistent, we call this task \textit{distributed detection} throughout this paper. The aim of the whole group is to detect the underlying true state of an object. To complete the task, a variety of distributed algorithms are designed in the literature to effectively aggregate information scattered all over the network.

In this paper, we focus on a class of distributed detection algorithms which involve the implementation of Bayes' law in a distributed setting. It is well known that the standard Bayes' law is very useful in detection, estimation, hypothesis testing, and other similar applications. By continuously observing new data or other useful information, an individual could eventually learn the true state. In a networked setting, it is a common phenomenon that any single agent could not learn the true state by itself. However, communicating with others in the network might bring the agents more useful information, and under certain conditions, all agents might eventually learn the true state collectively.

Our work in this paper is directly motivated by \cite{nedic,lalita,UCLA,jad_nb,tahbaz} where different but quite similar distributed detection algorithms are developed. All of these algorithms combine certain kinds of consensus protocols with Bayesian updates. They are different mainly in the aspect of the type of consensus protocol and the order of the two operations. In \cite{nedic}, agents first logarithmically aggregate their neighbors' beliefs to form a new prior belief, and then update their own beliefs using Bayes' law. In \cite{lalita}, in contrast to that in \cite{nedic}, an distributed detection algorithm with local Bayesian update first is proposed, i.e., the two operations change their order. In stead of logarithmically aggregating neighbors' posterior beliefs like that in \cite{lalita}, an algorithm with linearly aggregation is proposed in \cite{UCLA}. In \cite{jad_nb}, each agent first compute its Bayesian posterior belief based on its private observation, and then linearly combine it with neighbors' prior beliefs which can be interpreted as communication delay in the dynamics. The distributed detection rule in \cite{tahbaz}, also involving delayed communication, logarithmically combines the Bayesian posterior based on private observation and neighbors' prior beliefs, which is can be viewed as a logarithmic analog of the rule in \cite{jad_nb}.

In this paper, we first provide a systemic discussion about the class of distributed detection algorithms which combine certain kinds of consensus protocols with Bayesian updates, and show their essential similarities and differences. Some of the algorithms are proposed in the literature most recently, and others are first proposed in this paper. Then, we provide numerical examples to compare their performance in distributed detection problems. Some qualitative results are given which might provide us insight into the design of more efficient distributed detection algorithms.

This paper is organized as follows. In Sec. 2 we discuss and compare the distributed detection algorithms. In Sec. 3 we provide numerical examples to analyze the factors which influence the efficient of distributed detection algorithms. Concluding remarks are given in Sec. 4.

\section{Distributed Detection Algorithms Based on Consensus and Bayesian Updates}

\subsection{Preliminaries}

Consider a social network as a directed graph $G=(V,E)$, where
$V=\{1,2,\cdots,n\}$ is the node set and $E\subset V\times V$ is
the edge set. Each node in $V$ represents an agent, and the edge from $i$ to $j$, denoted by
the order pair $(i,j)\in E$ , captures the fact that agent $i$ is a
neighbor of agent $j$, and $j$ can receive some information from $i$. The
set of neighbors of agent $i$ is denoted by $N_i=\{j\in V:(j,i)\in
E\}$. Moreover, weight $a_{ij}\in [0,1]$ is assigned to any ordered pair of agents such that $a_{ij}>0$ if and only if $j\in N_i$. The weight $a_{ii}\ge 0$ is the self-weight of agent $i$, and we posit that $\sum_{i=1}^n a_{ij}=1$.

Let $\theta$ denote a state of the object we are concerned with, and all
the possible states compose a state set $\Theta=[\theta_1,\theta_2,\cdots,\theta_m]$, in which the true
state is denoted by $\theta ^ \ast $. From the point of view of agent $i$ at time $t$, the probability of state $\theta$ being true is denoted by $\mu_{i,t}(\theta)$, which is called the \textit{belief} of agent $i$ on $\theta$. Thus, agent $i$'s belief $\mu_{i,t}=[\mu_{i,t}(\theta_1), \mu_{i,t}(\theta_2), \cdots, \mu_{i,t}(\theta_m)]\in \mathbb{P}(\Theta)$
is a probability distribution over $\Theta$, where $\mathbb{P}(\Theta)$ is the set of all possible probability distribution over $\Theta$.

Conditional on the underlying true state, at each time period $t>0$,
a signal vector $s_t=(s_{1,t},s_{2,t},\cdots,s_{n,t})\in S$ is generated
according to the likelihood function $\ell (s_t|\theta ^\ast)$,
where $s_{i,t}$ is the signal observed by agent $i$ and $S$ is the
signal space. For each observed signal $s$ and each possible state
$\theta$, agent $i$ holds a corresponding private signal structure
$\ell_i(s|\theta)>0$, representing the probability that it believes
signal $s$ appears if the true state is $\theta$. We assume that the
private signal structure of agent $i$ about the true state $\theta ^
\ast$ is the $i$-th marginal of $\ell(\cdot |\theta^\ast)$, which
means the agent has a perfect prior information about the true
state. If there exists a state $\bar{\theta}\ne \theta ^\ast$
satisfying that $\ell_i(s|\bar{\theta})=\ell_i(s|\theta ^\ast)$ for
all signal $s$, we call $\bar{\theta}$ observationally equivalent to
the true state. That is to say, state $\bar{\theta}$ and the
underlying true state $\theta ^ \ast$ arouse exactly the same
signals according to the same probability in agent $i$'s eyes, and
thus, he cannot tell these two states apart only by observing the
signals. All the states that observationally equivalent to $\theta
^\ast$ from the point of view of agent $i$ compose a set $
\bar{\Theta }_i=\{ \theta \in {\Theta} : \ell_i (s|\theta) =
\ell_i (s|\theta ^\ast),~\forall ~s\in S\}$. If $\cap_{i=1}^n \bar{\Theta }_i=\{\theta^{\ast}\}$, we say the true state $\theta^\ast$ is globally identifiable.

In the next, we will describe six distributed detection algorithms used to detect the underlying true state, which all can be viewed as combinations of consensus protocols and Bayesian updates.

\subsection{Interpreting the Bayesian Posterior as the Solution of an Optimization Problem}

The standard Bayesian posterior obtain by agent $i$ based on its observation $s_{i,t+1}$ is as follows:
\begin{equation}
\label{eq1}
  \mu_{i,t+1}(\theta)=\frac{\mu_{i,t}(\theta)\ell_i(s_{i,t+1}|\theta)}{\sum_{\theta_k\in \Theta}\mu_{i,t}(\theta_k)\ell_i(s_{i,t+1}|\theta_k)}~, ~~~\theta\in \Theta.
\end{equation}

As point out in \cite{optimal1,optimal2,nedic}, the posterior belief can be interpreted as the solution of the following optimization problem
\begin{multline}
\label{eq2}
  \mu_{i,t+1}=\mathop{\arg\min}_{\pi\in \mathbb{P}(\Theta)}\Big\{D_{KL}(\pi\|\mu_{i,t}) \\ -\sum_{\theta_k\in \Theta}\pi(\theta_k)\log\left(\ell_i(s_{i,t+1}|\theta_k)\right)\Big\}
\end{multline}
where $D_{KL}(\pi\|\mu_{i,t})$ is the Kullback-Leibler divergence (the KL-divergence for short) between probability distributions $\pi$ and $\mu_{i,t}$ with the following definition
\begin{equation}
  D_{KL}(\pi\|\mu_{i,t})=\sum_{\theta_k\in\Theta}\pi(\theta_k)\log\frac{\pi(\theta_k)}{\mu_{i,t}(\theta_k)}. \nonumber
\end{equation}

Note that the first term on the right hand side of (\ref{eq2}) measures the difference between the distributions $\pi$ and $\mu_{i,k}$, and the second term is the maximum likelihood estimation given the observation $s_{i,t+1}$. Thus, the posterior distribution can be viewed as a tradeoff between the prior belief and the observation.

In a network setting, by introducing neighbors' prior distribution into the optimization problem (\ref{eq2}), we obtain the following new optimization problem
\begin{multline}
\label{eq3}
  \mu_{i,t+1}=\mathop{\arg\min}_{\pi\in \mathbb{P}(\Theta)}\bigg\{\sum_{j=1}^n a_{ij}D_{KL}(\pi\|\mu_{j,t}) \\ -\sum_{\theta_k\in \Theta}\pi(\theta_k)\log\left(\ell_i(s_{i,t+1}|\theta_k)\right)\bigg\}.
\end{multline}

Note that
\begin{eqnarray}
  && \!\!\!\!\!\!\!\!\sum_{j=1}^n a_{ij}D_{KL}(\pi\|\mu_{j,t}) \nonumber \\
  =&& \!\!\!\!\!\!\!\!\sum_{j=1}^n a_{ij}\sum_{\theta_k\in\Theta}\pi(\theta_k)\log\frac{\pi(\theta_k)}{\mu_{j,t}(\theta_k)} \nonumber \\
  =&& \!\!\!\!\!\!\!\!\sum_{\theta_k\in\Theta}\pi(\theta_k)\sum_{j=1}^n a_{ij}\log\frac{\pi(\theta_k)}{\mu_{j,t}(\theta_k)} \nonumber \\
  =&& \!\!\!\!\!\!\!\!\sum_{\theta_k\in\Theta}\pi(\theta_k)\log\prod_{j=1}^n\left(\frac{\pi(\theta_k)}{\mu_{j,t}(\theta_k)}\right)^{a_{ij}} \nonumber \\
  =&& \!\!\!\!\!\!\!\!\sum_{\theta_k\in\Theta}\pi(\theta_k)\log\frac{\pi(\theta_k)}{\prod_{j=1}^n \mu^{a_{ij}}_{j,t}(\theta_k)} \nonumber \\
  =&& \!\!\!\!\!\!\!\!D_{KL}\bigg(\pi\|\prod_{j=1}^n\mu^{a_{ij}}_{j,t}\bigg) \nonumber
\end{eqnarray}
which is the KL-divergence between $\pi$ and the geometric mean of the prior beliefs of agent $i$ and its neighbors.

By the above derivation, the optimization problem (\ref{eq3}) can be rewritten as
\begin{multline}
\label{eq4}
  \mu_{i,t+1}=\mathop{\arg\min}_{\pi\in \mathbb{P}(\Theta)}\bigg\{D_{KL}\Big(\pi\|\prod_{j=1}^n\mu^{a_{ij}}_{j,t}\Big) \\ -\sum_{\theta_k\in \Theta}\pi(\theta_k)\log\left(\ell_i(s_{i,k+1}|\theta_k)\right)\bigg\}.
\end{multline}
The solution of (\ref{eq3}) (also (\ref{eq4})), is the Bayesian posterior belief corresponding to the prior $\prod_{j=1}^n\mu^{a_{ij}}_{j,t}$, which has the following form
\begin{equation}
\label{eq5}
  \mu_{i,t+1}(\theta)=\frac{\prod_{j=1}^n\mu^{a_{ij}}_{j,t}(\theta)\ell_i(s_{i,t+1}|\theta)}{\sum_{\theta_k\in \Theta}\prod_{j=1}^n\mu^{a_{ij}}_{j,t}(\theta_k)\ell_i(s_{i,t+1}|\theta_k)}, ~\theta \in \Theta.
\end{equation}
The updating rule (\ref{eq5}) is a distributed algorithm whereby each agent first aggregates the beliefs of its neighbors and itself as a new prior belief via weighted geometric average, and then uses Bayes' law to compute posterior distribution. For simplicity, in this paper we call the rule (\ref{eq5}) \textbf{\textit{LoAB}} (Logarithmic Aggregation and Bayesian update). The rule \textit{LoAB} has been originally proposed and studied by Nedi\'c \textit{et al.} in \cite{nedic}, and the sufficient condition under which agents can learn the underlying true state using (\ref{eq5}) is summarized as follows: \\
\textit{
\textbf{Condition 1:}
\begin{enumerate}
\renewcommand{\labelenumi}{(\theenumi)}
\item The time-varying network is B-strongly connected, i.e., there is an integer $B\ge 1$ such that the network is jointly strongly connected across every $B$ time slots.
\item Any positive weight has a constant lower bound $\eta>0$, i.e., if $a_{ij}>0$ then $a_{ij}\ge\eta$ ~for all $i,j\in V$.
\item All agents have positive self-weights, i.e., $a_{ii}>0$ for all $i\in V$.
\item All agents have positive initial belief on $\theta^{\ast}$.
\item The true state $\theta^\ast$ is globally identifiable.
\end{enumerate}}

In \cite{nedic}, Nedi\'c \textit{et al.} further consider a more general case where the true state $\theta^{\ast}$ might not be listed as one of the possible states. They define a set of state $\Omega_i$ for each agent $i$, where $\Omega_i=\mathop{\arg\min}_{\theta_k\in \Theta}D_{KL}(\ell_i(\cdot|\theta^{\ast})|\ell_i(\cdot|\theta_k))$, and let $\Omega^{\ast}\triangleq\cap_{i=1}^n\Omega_i$ which is not empty. That is to say, even though the true state might not be considered as a possible state, there exist some states which best explain the observations from the point of view of all agents. They prove that under such a relaxed condition, $\mu_{i,t}(\theta)\rightarrow 0$ almost surely as $t\rightarrow \infty$ for all $i$ and $\theta\notin \Omega^{\ast}$.   This implies that if there is only one state in $\Omega^{\ast}$, all agents eventually assign belief of one on this state which is closest to the true state.

Next we propose an alternative way to introduce neighbors' information into the Bayesian update (\ref{eq2}): instead of computing the weighted average of all KL-divergences between $\pi$ and prior beliefs like that in (\ref{eq3}), we can compute the KL-divergence between $\pi$ and the weighted average of neighbors' prior beliefs. The new optimization problem is as follows
\begin{multline}
\label{eq6}
  \mu_{i,t+1}=\mathop{\arg\min}_{\pi\in \mathbb{P}(\Theta)}\bigg\{D_{KL}\Big(\pi\|\sum_{j=1}^n a_{ij}\mu_{j,t}\Big) \\ -\sum_{\theta_k\in \Theta}\pi(\theta_k)\log\left(\ell_i(s_{i,k+1}|\theta_k)\right)\bigg\}.
\end{multline}
The solution of (\ref{eq6}) is the Bayesian posterior distribution corresponding to the prior $\sum_{j=1}^n a_{ij}\mu_{j,t}$, i.e., for any $\theta\in \Theta$
\begin{equation}
\label{eq7}
  \mu_{i,t+1}(\theta)=\frac{\sum_{j=1}^n a_{ij}\mu_{j,t}(\theta)\ell_i(s_{i,t+1}|\theta)}{\sum_{\theta_k\in \Theta}\Big(\sum_{j=1}^n a_{ij}\mu_{j,t}(\theta_k)\ell_i(s_{i,t+1}|\theta_k)\Big)}.
\end{equation}
The essential difference between (\ref{eq5}) and (\ref{eq7}) is that the former aggregates prior beliefs via geometric average while the latter via arithmetics average. Here we call the algorithm (\ref{eq7}) \textbf{\textit{LiAB}} (Linear Aggregation and Bayesian update). We simply propose this algorithm in this paper without strict theoretical analysis. To the best of our knowledge, this algorithm has not been proposed and studied in the existing papers. Therefore, theoretically analyzing its performance is still an open question.

\subsection{Aggregation of Bayesian Posterior Distributions}

 A common feature of the algorithms \textit{LiAB} and \textit{LoAB} is that they first aggregate local prior beliefs linearly or logarithmically as a new prior and then use the Bayes' law to compute the posterior distribution. We might change the order of the two steps and obtain two new algorithms.

If we let each agent first update its belief distribution based on its private observation and then exchange the posterior distribution with its neighbors via weighted geometric average, we obtain the following algorithm which is called \textbf{\textit{BLoA}}(Bayesian update and Logarithmic Aggregation)
\begin{equation}
\label{eq8}
\mu_{i,t+1}(\theta)=\frac{\prod_{j=1}^n\tilde{\mu}^{a_{ij}}_{j,t+1}(\theta)}{\sum_{\theta_k\in \Theta}\prod_{j=1}^n\tilde{\mu}^{a_{ij}}_{j,t+1}(\theta_p)}
  \end{equation}
where
\begin{equation}
\label{eq9}
   \tilde{\mu}_{i,t+1}(\theta)=\frac{\mu_{i,t}(\theta)\ell_i(s_{i,t+1}|\theta)}{\sum_{\theta_k\in \Theta}\mu_{i,t}(\theta_k)\ell_i(s_{i,t+1}|\theta_k)}.
  \end{equation}
The denominator in (\ref{eq8}) is added to ensure that $\mu_{i,t+1}$ is still a well-defined probability distribution over $\Theta$.

The rule (\ref{eq8}) has been proposed and extensively studied in \cite{lalita}. The sufficient condition for detecting the true state is summarized as follows:\\
\textit{
\textbf{Condition 2:}
\begin{enumerate}
\renewcommand{\labelenumi}{(\theenumi)}
\item The network is strongly connected.
\item All agents have positive initial belief on all $\theta \in \Theta$.
\item For every pair $\theta_p\ne \theta_q$, there is at least one agent $i$ for which the KL-divergence $D_{KL}(\ell_i(\cdot|\theta_p)\|\ell_i(\cdot|\theta_q))>0$.
\end{enumerate}}
Compared with \textit{Condition 1}, the terms in \textit{Condition 2} are more stringent. The second term requires that not only the initial beliefs of all agents on the true state are positive, but also beliefs on all other states must be positive. The third term implies that there is no state that is observationally equivalent to any other state from the point of view of all agents in the network, i.e., all states are globally identifiable.

Note that the rule in \cite{lalita} is not in the form like (\ref{eq8}) but in the following form
\begin{equation}
\label{eq10}
\mu_{i,t+1}(\theta)=\frac{\exp\Big(\sum_{j=1}^n a_{ij}\log \tilde{\mu}_{j,t+1}(\theta)\Big)}{\sum_{\theta_k\in \Theta}\exp\Big(\sum_{j=1}^n a_{ij}\log \tilde{\mu}_{j,t+1}(\theta_k)\Big)}.
\end{equation}
In fact, the rule (\ref{eq10}) is identical to (\ref{eq8}) since
\begin{eqnarray}
&&\!\!\!\!\!\!\!\!\frac{\exp\Big(\sum_{j=1}^n a_{ij}\log \tilde{\mu}_{j,t+1}(\theta)\Big)}{\sum_{\theta_k\in \Theta}\exp\Big(\sum_{j=1}^n a_{ij}\log \tilde{\mu}_{j,t+1}(\theta_k)\Big)} \nonumber \\
=&&\!\!\!\!\!\!\!\!\frac{\exp\Big(\sum_{j=1}^n \log \tilde{\mu}^{a_{ij}}_{j,t+1}(\theta)\Big)}{\sum_{\theta_k\in \Theta}\exp\Big(\sum_{j=1}^n \log \tilde{\mu}^{a_{ij}}_{j,t+1}(\theta_k)\Big)} \nonumber \\
=&&\!\!\!\!\!\!\!\!\frac{\exp\Big(\log \prod_{j=1}^n \tilde{\mu}^{a_{ij}}_{j,t+1}(\theta)\Big)}{\sum_{\theta_k\in \Theta}\exp\Big(\log \prod_{j=1}^n \tilde{\mu}^{a_{ij}}_{j,t+1}(\theta_k)\Big)} \nonumber \\
=&&\!\!\!\!\!\!\!\!\frac{\prod_{j=1}^n \tilde{\mu}^{a_{ij}}_{j,t+1}(\theta)}{\sum_{\theta_k\in \Theta} \prod_{j=1}^n \tilde{\mu}^{a_{ij}}_{j,t+1}(\theta_k)}. \nonumber
\end{eqnarray}
From the above derivation it is not hard to understand why we call rules (\ref{eq5}) and (\ref{eq8}) logarithmic aggregations, since both of them involve geometric averages which can be written in logarithmic forms like that in (\ref{eq10}).

Similar to (\ref{eq8}) in the sense of aggregating Bayesian posterior beliefs of neighbors, an algorithm with geometric average being replaced by arithmetic average is proposed and extensively studied in \cite{UCLA}, which is called \textbf{\textit{BLiA}}(Bayesian update and Linear Aggregation) here
\begin{equation}
\label{eq11}
\mu_{i,t+1}(\theta)=\sum_{j=1}^n a_{ij}\tilde{\mu}_{j,t+1}(\theta)
  \end{equation}
where $\tilde{\mu}_{j,t+1}(\theta)$ is identical to that in (\ref{eq9}). The sufficient conditions under which agents can detect the true state can be summarized as follows:\\
\textit{
\textbf{Condition 3:}
\begin{enumerate}
\renewcommand{\labelenumi}{(\theenumi)}
\item The weight matrix $A=[a_{ij}]$ is primitive.
\item There exists at least one agent with positive initial belief on the true state.
\item For each agent $i$, there exists at least one prevailing signal $s_i^o$ such that $\ell_i(s_i^o|\theta ^\ast) - \ell_i(s_i^o|\theta)\ge \delta_k^o>0$ for any $\theta\notin \bar{\Theta }_i$.
\end{enumerate}}

Compared with \textit{LoAB} and \textit{BLoA}, the algorithm \textit{BLiA} requires more relaxed conditions in some aspects to detect the underly true state. For instance, it only needs at least one agent having positive initial belief on the true state. And also, the requirement of primitive matrix is more relaxed than that with positive diagonal elements (i.e., positive self-weights).

\subsection{Aggregation of Personal Bayesian Posterior and Others' Prior}

The following two algorithms, like \textit{BLoA} and \textit{BLiA}, also contain personal Bayesian update and communication with neighbors. However, agents exchange prior distributions with their neighbors rather than posterior distributions. We may consider that this sort of algorithms introduce communication delay into the dynamics such that agents could not receive their neighbors' latest information (the posteriors) but only delayed information (the priors before Bayesian update).

There are also two ways to aggregate neighbors' information. If the agent aggregate its personal posterior and its neighbors' priors via arithmetic average, we have the following algorithm called \textbf{\textit{BLiAD}} (Bayesian update and Linear Aggregation of Delayed information)
\begin{equation}
\label{eq12}
\mu_{i,t+1}(\theta)=a_{ii}\tilde{\mu}_{i,t+1}(\theta)+\sum_{j\in N_i}a_{ij}\mu_{j,t}(\theta).
  \end{equation}

The rule (\ref{eq12}) is originally proposed in \cite{jad_nb} in the context of social learning. The sufficient condition under which agents can learn the underlying true state using (\ref{eq12}) can be summarized as follows: \\
\textit{
\textbf{Condition 4:}
\begin{enumerate}
\renewcommand{\labelenumi}{(\theenumi)}
\item The network is strongly connected.
\item All agents have strictly positive self-weights, i.e., $a_{ii}>0$ for all $i\in V$.
\item There exists at least one agent with positive initial belief on the true state $\theta^{\ast}$.
\item The true state $\theta^\ast$ is globally identifiable.
\end{enumerate}}

Replacing the arithmetic average in \textit{BLiAD} by the geometric average, we obtain another algorithm here called \textbf{\textit{BLoAD}} (Bayesian update and Logarithmic Aggregation of Delayed information) as follows
  \begin{equation}
  \label{eq13}
\mu_{i,t+1}(\theta)=\frac{\tilde{\mu}^{a_{ii}}_{i,t+1}(\theta)\prod_{j\in N_i}\mu^{a_{ij}}_{j,t}(\theta)}{\sum_{\theta_k\in \Theta}\left(\tilde{\mu}^{a_{ii}}_{i,t+1}(\theta_k)\prod_{j\in N_i}\mu^{a_{ij}}_{j,t}(\theta_k)\right)}.
  \end{equation}

Similar to (\ref{eq13}), in \cite{tahbaz} Rad and Tahbaz-Salehi propose a distributed estimation algorithm as follows:
  \begin{equation}
  \label{eq14}
\log\mu_{i,t+1}(\theta)=\lambda_i\log\ell_i(s_{i,t+1}|\theta)+\sum_{j=1}^n a_{ij}\log\mu_{j,t}(\theta)+c_{i,t}
  \end{equation}
where $\lambda_i>0$ is the weight that agent $i$ assigns to its private observations and $c_{i,t}$ is a normalization constant, not dependent on $\theta$, which ensures that $\mu_{i,t+1}$ is a well-defined probability distribution over $\Theta$.

The rule (\ref{eq14}) is identical to (\ref{eq13}) by choosing the following values for the parameters $\lambda_i$ and $c_{i,t}$:
\begin{equation}
  \lambda_i=a_{ii} \nonumber
\end{equation}
and
\begin{multline}
  c_{i,t}=-\log\Bigg[\bigg(\sum_{\theta_k\in \Theta}\mu_{i,t}(\theta_k)\ell_i(s_{i,t+1}|\theta_k)\bigg)^{a_{ii}}\times \\ \sum_{\theta_k\in \Theta}\Big(\tilde{\mu}^{a_{ii}}_{i,t+1}(\theta_k)\prod_{j\in N_i}\mu^{a_{ij}}_{j,t}(\theta_k)\Big)\Bigg]. \nonumber
\end{multline}
In fact, the rule (\ref{eq14}) can be written in the following form:
  \begin{eqnarray}
  \label{eq15}
&&\!\!\!\!\!\!\!\!\mu_{i,t+1}(\theta) \nonumber \\
=&&\!\!\!\!\!\!\!\!\exp\Big(\lambda_i\log\ell_i(s_{i,t+1}|\theta)+\sum_{j=1}^n a_{ij}\log\mu_{j,t}(\theta)+c_{i,t}\Big)\nonumber \\
=&&\!\!\!\!\!\!\!\!\exp\Big(\log\ell^{\lambda_i}_i(s_{i,t+1}|\theta)+\log\prod_{j=1}^n\mu^{a_{ij}}_{j,t}(\theta)+c_{i,t}\Big)\nonumber \\
=&&\!\!\!\!\!\!\!\!\ell^{\lambda_i}_i(s_{i,t+1}|\theta)\prod_{j=1}^n\mu^{a_{ij}}_{j,t}(\theta)\cdot e^{c_{i,t}} \nonumber \\
=&&\!\!\!\!\!\!\!\!\ell^{a_{ii}}_i(s_{i,t+1}|\theta)\mu^{a_{ii}}_{i,t}(\theta)\prod_{j\in N_i}\mu^{a_{ij}}_{j,t}(\theta)\cdot e^{c_{i,t}} \nonumber \\
=&&\!\!\!\!\!\!\!\!\frac{\tilde{\mu}^{a_{ii}}_{i,t+1}(\theta)\prod_{j\in N_i}\mu^{a_{ij}}_{j,t}(\theta)}{\sum_{\theta_k\in \Theta}\left(\tilde{\mu}^{a_{ii}}_{i,t+1}(\theta_k)\prod_{j\in N_i}\mu^{a_{ij}}_{j,t}(\theta_k)\right)}.
  \end{eqnarray}

The rule (\ref{eq14}) has been theoretically studied in \cite{tahbaz}, and the sufficient condition for detecting the true state is summarized as follows:\\
\textit{
\textbf{Condition 5:}
\begin{enumerate}
\renewcommand{\labelenumi}{(\theenumi)}
\item The network is strongly connected.
\item All agents have positive initial prior belief on all $\theta \in \Theta$.
\item The true state $\theta^\ast$ is globally identifiable.
\end{enumerate}}

\begin{figure*}[!b]
  \centering
  \includegraphics[width=0.9\hsize]{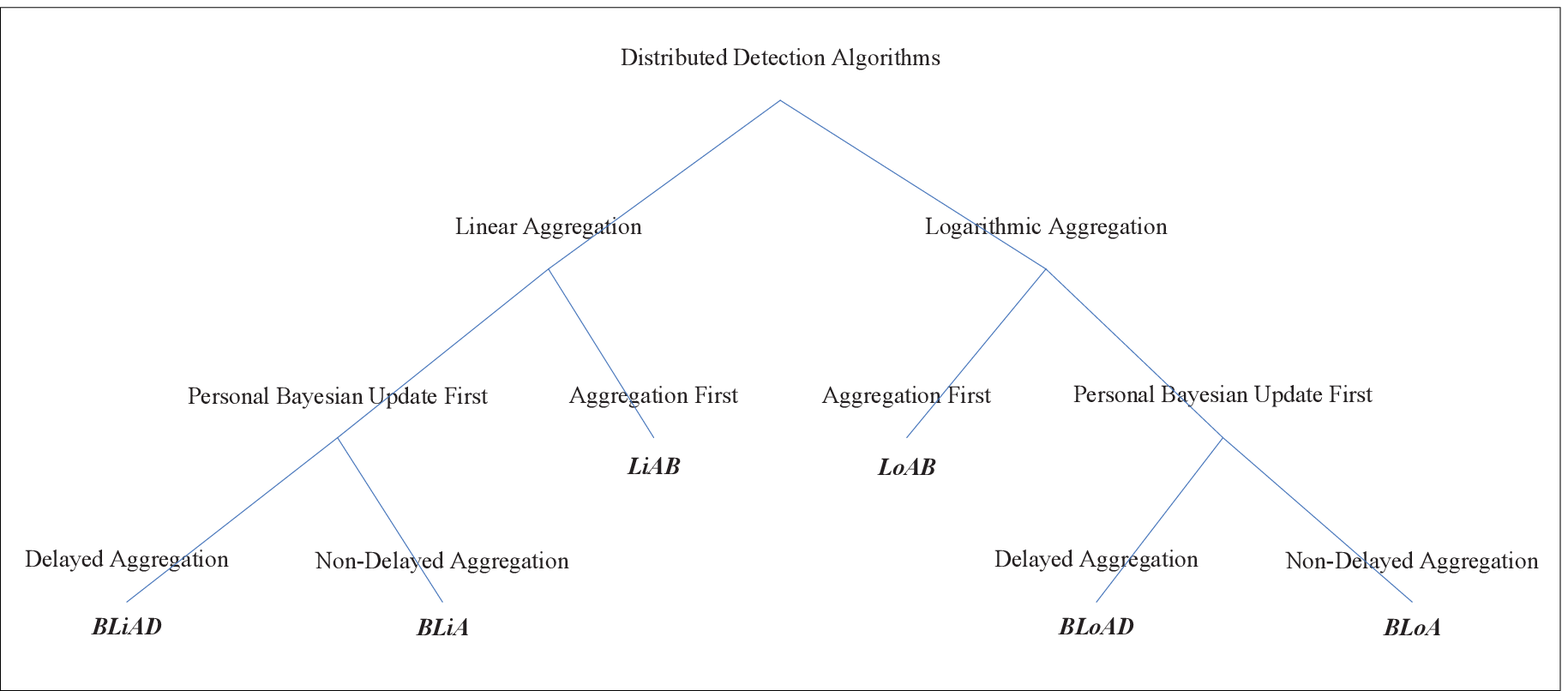}
  \caption{Classification tree of the six distributed detection algorithms}
  \label{fig1}
\end{figure*}

As a summarization, we provide the classification tree of the six distributed detection algorithms in Fig.~\ref{fig1}.

\subsection{Comparison with the Bayesian Update of a Single Agent}

L.~J.~Savage has pointed out in \cite{savage} that, barring two banal exceptions, a single agent
becomes almost certain of the true state when the amount of its
observation increases infinitely. One exception is that the initial
belief of the true state is zero. This is very easy to understand.
If the belief is zero, then, no matter what signal is observed, the
posterior belief of the true state is still zero. The other
exception occurs when there exists a state which arouses exactly the
same signals as the true state does, i.e., observationally
equivalent state exists.

When an agent is situated in a network setting, the above two requirements might be relaxed. For instance, it has been proven that the rules \textit{LoAB, BLoA, BLiA, BLiAD,} and \textit{BLoAD} only require the true state being globally identifiable. We conjecture that \textit{LiAB} might also work well with the same condition, even though theoretical analyses are not available yet.

It is not hard to see that all rules involving geometric averages, such as \textit{LoAB, BLoA,} and \textit{BLoAD}, still need the requirement of non-zero initial beliefs of all agents on the true state. For the rules \textit{BLiA} and \textit{BLiAD} which contains linear aggregations, it has been proven in \cite{UCLA} and \cite{jad_nb}, respectively, that at least one agent with non-zero initial belief on the true state is enough for a correct detection.

For the rules involving delayed information such as \textit{BLiAD} and \textit{BLoAD}, the requirement of non-zeros self-weights must be satisfied, at least for part of the agents. Because if all self-weights are zeros, any new observation will be discarded and $\textit{BLiAD}$ and $\textit{BLoAD}$ specialize to traditional consensus protocols with arithmetic average and geometric average, respectively.

\section{Numerical Examples}

In this section, we exam the effectiveness of each distributed detection rule by numerical examples. Our test platform is a nearest-neighbor coupled network, which might represent a sensor network or a robot network where, restricted by its communication range, each agent can only exchange information with a given number of its closest neighbors. The following Fig.~\ref{fig2} shows a schematic diagram of a nearest-neighbors coupled network of 20 agents, where each agent could only interact with 5 closest agents (including itself) and all of the edges are bi-directed.
\begin{figure}[!htb]
  \centering
  \includegraphics[width=0.5\hsize]{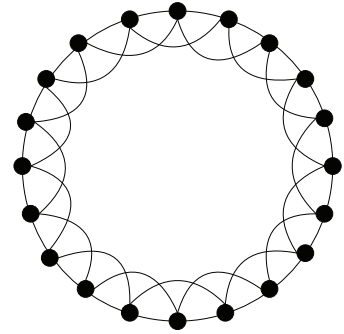}
  \caption{A nearest-neighbor coupled network of 20 agents}
  \label{fig2}
\end{figure}

Simulations are performed on three possible states, i.e.,
$\Theta=\{\theta_1,\theta_2,\theta_3\}$ in which $\theta_3$ is set to be the true state. Any agent $i$'s belief distribution at
time $t$ is $\mu_{i,t}=[\mu_{i,t}(\theta_1),\mu_{i,t}(\theta_2),\mu_{i,t}(\theta_2)]\in
[0,1]^3$. The initial belief is uniformly distributed in the interval
[0,1] and subject to
$\sum_{k=1}^3\mu_{i,t}(\theta_k)=1$.

The signals generated by the true state are $\{s_1,s_2\}$. We assume
that signal $s_1$ appears with possibility of 0.8, and $s_2$ with 0.2,
which implies the private signal structure about $\theta_3$ of any
agent $i$ should be $\ell_i(s_1|\theta_3)=0.8$ and
$\ell_i(s_2|\theta_3)=0.2$.

Let half of agents, denoted by $V_1\subset V$, have the signal structures
$\ell_i(s_1|\theta_1)=0.8$, $\ell_i(s_2|\theta_1)=0.2, \ell_i(s_1|\theta_2)=0.5$, and $\ell_i(s_2|\theta_2)=0.5$ ($i\in V_1$). That is to say, the states $\theta_1$ and $\theta_3$ are equivalent to agents in $V_1$. Let the other half of agents, denoted by $V_2=V\backslash V_1$, have the following signal structures: $\ell_i(s_1|\theta_1)=0.2$, $\ell_i(s_2|\theta_1)=0.8, \ell_i(s_1|\theta_2)=0.8$, and $\ell_i(s_2|\theta_2)=0.2$ ($i\in V_2$), i.e., the state $\theta_2$ and $\theta_3$ are observationally equivalent to agents in $V_2$. The true state is unidentifiable to any single agent, but globally identifiable.

In our first simulation, we let agents from $V_1$ be close to each other, and the same to $V_2$, i.e., agents belong to the same set form a cluster. In the second simulation, we mix all agents in the sense that each agent from $V_1$ is located between two agents from $V_2$, and vice versa.

\begin{figure}[!htb]
  \centering
  \includegraphics[width=\hsize]{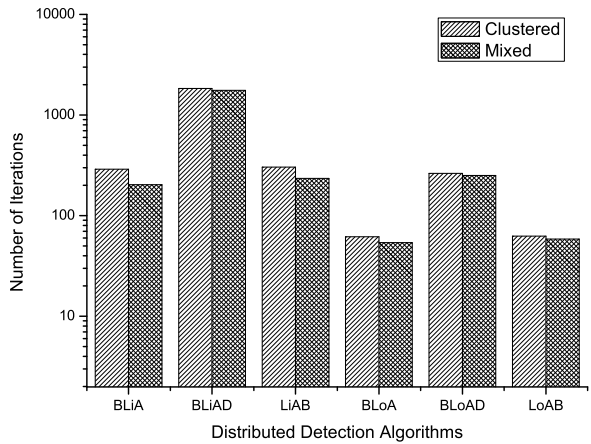}
  \caption{Comparison of six distributed detection algorithms}
  \label{fig3}
\end{figure}

We focus on the number of iterations of update for each algorithm to detect the true state. If for all $i\in V$, $|\mu_{i,t}(\theta_3)-1|\le 10^{-3}$, we say the whole group collectively detect the true state. The result is shown in Fig.~\ref{fig3}, which is the average of 100 realizations.
From Fig.~\ref{fig3}, we have the following observations:
\begin{enumerate}
\renewcommand{\labelenumi}{(\theenumi)}
\item All of the six distributed detection algorithms are effective in detecting the true state.
\item If other aspects are identical, aggregating information via geometric average (i.e., \textit{BLoA, BLoAD}, and \textit{LoAB}) is much faster than that via arithmetic average (i.e., \textit{BLiA, BLiAD}, and \textit{LiAB}).
\item Using delayed information in aggregation (i.e., \textit{BLiAD} and \textit{BLoAD}) makes the detection much slower.
\item Bayesian update first or aggregating information from neighbors first does not influence the efficiency of detection apparently.
\item Mixing agents with different signal structures promotes the rate of detection.
\end{enumerate}

The above items (2) and (3) are in accordance with the theoretical result in \cite{lalita} where \textit{BLoA} is compared with \textit{BLiAD}. One of the main results is that, the lower bound on the rate of detection by using \textit{BLoA} is even greater than the upper bound of \textit{BLiAD}. The item (3) is also in accordance with the theoretical result in \cite{UCLA}, in which \textit{BLiA} is compared with \textit{BLiAD}, and the former is much faster in detecting the true state.

\section{Concluding Remarks}

In this paper, we discuss a class of distributed detection algorithms in which personal Bayesian updates are combined with some types of consensus protocols. We focus on six algorithms and classify them according to the type of consensus protocol, the order of Bayesian update and consensus, and whether time-delayed information is involved in the interaction. By comparison, we have a systematic impression of these distributed detection algorithms, which might lead us to establishing more refined conditions under which agents could detect the underlying true state in distributed settings. For instance, could the terms (2) and (3) in \textit{Condition 2} be replaced by more relaxed requirements, say the terms (4) and (5) in \textit{Condition 1}, respectively? And also, could the \textit{Conditions 2, 3, 4,} and \textit{5} be relaxed to time-varying networks like that in \textit{Condition 1}? Through numeric examples, we obtain some qualitative results about the efficiency of these distributed algorithms, which might shed light on designing more efficient distributed detection algorithms in the future.

There are many other types of distributed detection algorithms proposed in the literature. For instance, instead of exchanging beliefs, agents can share their signal structures with their neighbors which could also result in a correct detection of the true state \cite{jad_detect,rao,olfati}. Also, Bayesian update is not the only choice in detection problem. More alternatives can be found in the literature (e.g., \cite{sensor1,sensor2,sensor3,Sayed}).

\balance

\end{document}